\documentclass{article}

\usepackage{spconf}
\usepackage[edges]{forest}
\usepackage{amsmath,amssymb,amsfonts}
\usepackage{algpseudocode}
\usepackage[ruled,linesnumbered]{algorithm2e}
\usepackage{graphicx}
\usepackage{textcomp}
\usepackage{xcolor}
\usepackage{comment}
\usepackage{listings}

\usepackage [english]{babel}
\usepackage [autostyle, english = american]{csquotes}
\MakeOuterQuote{"}

\usepackage[utf8]{inputenc}

\usepackage[a4paper,margin=2cm]{geometry}
\usepackage{tikz}
\usetikzlibrary{trees,positioning,shapes,shadows,arrows}

\usepackage{hyperref}
\usepackage{booktabs}
\usepackage{multirow}

\def\BibTeX{{\rm B\kern-.05em{\sc i\kern-.025em b}\kern-.08em
    T\kern-.1667em\lower.7ex\hbox{E}\kern-.125emX}}

\title{jazznet: A Dataset of Fundamental Piano Patterns for Music Audio Machine Learning Research}

\name{Tosiron Adegbija}
\address{Department of Electrical and Computer Engineering, University of Arizona, Tucson AZ, USA\\tosiron@arizona.edu}

\begin{document}

\maketitle

\begin{abstract}
This paper introduces the \textit{jazznet Dataset}, a dataset of fundamental jazz piano music patterns for developing machine learning (ML) algorithms in music information retrieval (MIR). The dataset contains 162520 labeled piano patterns, including chords, arpeggios, scales, and chord progressions with their inversions, resulting in more than 26k hours of audio and a total size of 95GB. The paper explains the dataset's composition, creation, and generation, and presents an open-source \textit{Pattern Generator} using a method called \textit{Distance-Based Pattern Structures (DBPS)}, which allows researchers to easily generate new piano patterns simply by defining the distances between pitches within the musical patterns. We demonstrate that the dataset can help researchers benchmark new models for challenging MIR tasks, using a convolutional recurrent neural network (CRNN) and a deep convolutional neural network. The dataset and code are available via: \\ \url{https://github.com/tosiron/jazznet}. %Finally, we summarize different possible uses of the dataset and report the performance of simple experiment using a \textit{convolutional recurrent neural network (CRNN)} model to predict different classes and sub-classes of musical patterns. 

\end{abstract}


\section{Introduction}

One of the most important and basic needs of machine learning (ML) research in music is the availability of free datasets. Unfortunately, the field of music has lagged behind other fields (like image and speech recognition) in the availability of high-quality datasets. This paper contributes to the growing body of available music audio datasets by presenting the \textit{jazznet dataset}, an extensible dataset of fundamental piano patterns.

There are notable efforts toward creating datasets for MIR, and jazznet is complementary to these efforts. Some existing music datasets include GTZAN \cite{tzanetakis02_gtzan}, MSD \cite{bertin11_MSD}, AudioSet \cite{gemmeke17_audioSet}, FMA dataset \cite{defferrard16_fmaDataset}, MusicNet \cite{thickstun16_musicnet}, and RWC \cite{goto03_rwcMusicDatabase}. The main difference between the jazznet dataset and the majority of existing datasets lies in the approach taken to creating the dataset: an emphasis on the fundamental patterns, rather than complete music pieces. This approach is inspired by how humans effectively learn piano music. Suppose you wanted to learn to play jazz piano; you would learn much faster and more effectively if you first understood the fundamental patterns---\textit{chords, arpeggios, scales,} and \textit{chord progressions}---on which full music pieces are based \cite{levine11_jazzPianoBook}. We focus on jazz piano music patterns to make the dataset creation tractable and because of the versatility of jazz music---the patterns in jazz music encompass several other musical genres (e.g., blues, country, pop, etc.). Given that an important hallmark of jazz music is the variety of expression, the dataset also contains the different forms---or \textit{inversions}---in which the patterns can be played.

The multi-class characteristic of the dataset also aims to model the hierarchy of difficulty humans experience in learning jazz music. For instance, it is easy to differentiate between the \textit{types} of patterns (e.g., chords vs. arpeggios vs. scales). Within each of the classes, it is substantially more difficult to recognize different \textit{modes} of patterns (e.g., major vs. minor or augmented vs. diminished chords). Finally, only a uniquely talented musician with \textit{absolute pitch} \cite{deutsch06_absolutePitchEnigma} can recognize specific modes (e.g., A major vs. C major).

In summary, we make two major contributions. First, we present the jazznet dataset\footnote{\url{https://doi.org/10.5281/zenodo.7192653}}, which we have curated based on a wide-ranging survey of jazz education resources and jazz standards. The dataset contains 162520 automatically generated and labeled piano music patterns, which results in 95GB and over 26k hours of audio. The primary objective of the jazznet dataset is to facilitate the development of ML models for challenging MIR tasks. Second, we have developed a \textit{Pattern Generator} that uses \textit{Distance-Based Pattern Structures (DBPS)} to facilitate the generation of piano patterns based on the distance between pitches within the patterns. This approach enables the easy generation of new piano patterns for ML research, and we have open-sourced the Python scripts\footnote{\url{https://github.com/tosiron/jazznet/tree/main/PatternGenerator}} that implement it for the convenience of users. %In what follows, we describe the dataset, the creation approach, the pattern generator, and highlight some possible uses of the dataset. %We also describe an experiment that uses a simple convolutional recurrent neural network (CRNN) model to predict different classes and sub-classes in multiple subsets of the dataset. The performance of the model is contextualized by comparing it to the performance of a jazz musician in recognizing the classes/sub-classes by ear.

\section{The Jazznet Dataset}

Although complementary to existing music datasets, jazznet is unique in several ways. Structurally, the most related datasets to jazznet are the NSynth \cite{engel17_nsynth} and MAPS \cite{emiya10_maps_dataset} datasets, with some notable differences. NSynth is created for exploring neural audio synthesis of musical notes with different pitches, timbres, and envelopes. However, it does not contain any musical patterns, only single notes, and although it represents more musical instruments and dynamics than jazznet, it is not comparable in terms of the variety of musical patterns contained in jazznet. On the other hand, the MAPS dataset is frequently used for automatic music transcription \cite{benetos18_musicTranscription} and contains isolated notes, chords, and music pieces. However, it contains fewer chords than jazznet and does not include scales, arpeggios, or chord progressions. Additionally, the MAPS dataset is significantly smaller than jazznet, with 31GB of data compared to 95GB. Most importantly, unlike previous datasets, jazznet is accompanied by piano pattern generators that enable researchers to extend the dataset to include new patterns.

While the dataset is created to be easily extensible, we curated the patterns for the current version by extensively surveying several resources for jazz music performance and education, such as The Jazz Piano Book \cite{levine11_jazzPianoBook}, The Jazz Theory Book \cite{levine11_jazzTheoryBook}), and more than 50 jazz standards \cite{gioia21_jazzStandards,hal-leonard17_jazzStandards}. We also interacted with several jazz musicians to develop a consensus on some of the common fundamental patterns, particularly chord progressions. However, we acknowledge that given the complexity and diversity of jazz music, the dataset does not cover all jazz piano patterns. Nevertheless, the provided pattern generator (Section \ref{sec:patternGen}) enables users to generate new patterns not in the dataset. In the following sections, we provide a background on music theory necessary for understanding the dataset's creation, describe the data generation approach, and then present the jazznet dataset, its features, the pattern generator, and potential applications.

\subsection{Brief background basic music theory}

The dataset is based on the notes of a standard 8-octave (88-key) piano. The main notes of a piano (and most other stringed instruments) are typically represented using 7 letters: $A, B, C, D, E, F, G$ (which correspond to the white keys on a piano). The notes in between these notes (i.e., the black keys) are described as \textit{sharp ($\sharp$)} to the immediately preceding note, or \textit{flat ($\flat$)} to the immediately succeeding note. For example, the note between C and D is referred to as C$\sharp$ or D$\flat$ depending on the context. An \textit{octave} is the interval between one musical pitch and another half or double its frequency. For example, the middle C on the piano (also called C4) has a frequency of 261.63 Hz, the next C (C5) has a frequency of 523.25 Hz, and the previous C (C3) has a frequency of 130.81 Hz. There are 12 `half steps' (also called \textit{semitones}) or 6 `full steps' (also called \textit{tones}) between each note and its octave. 

\begin{figure}[t]
		\centering
		\includegraphics[width=0.6\linewidth]{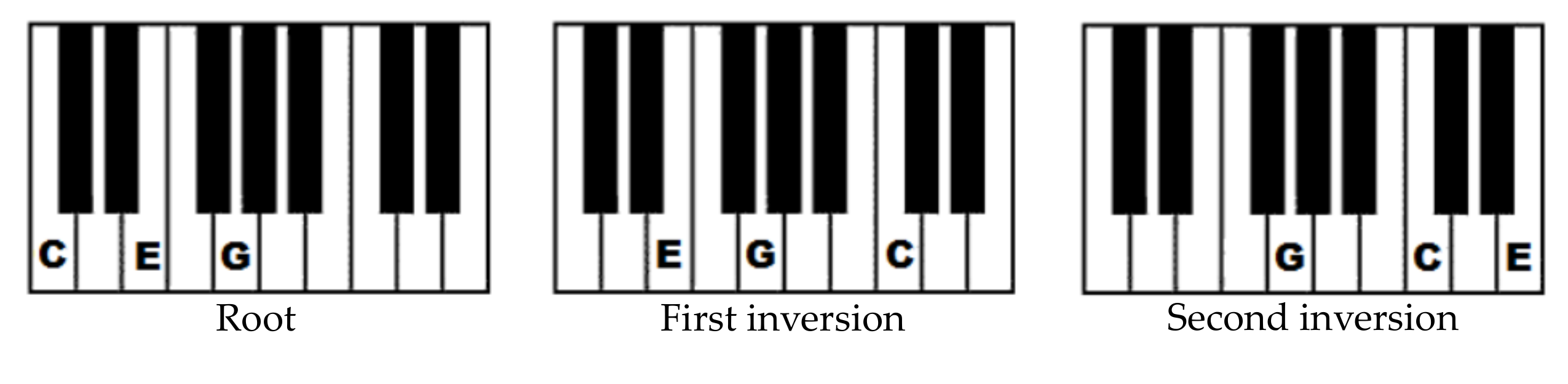}
		\vspace{-15pt}
		\caption{Illustration of the C major chord and its three forms: root, first, and second inversions}
		\label{fig:inversions}
       \vspace{-10pt}
\end{figure}

The dataset consists of four types of piano patterns: \textit{chords, arpeggios, scales} and \textit{chord progressions}. A chord is a harmonic set of pitches or frequencies consisting of multiple notes that are played or heard simultaneously. Chords can be categorized by the number of notes they contain: \textit{dyads} (2-note), \textit{triad} (3-note), \textit{tetrad} (4-note), etc. An arpeggio is a chord in which the notes are played one after another. A scale is a set of notes ordered by the frequency of the notes. The notes in these patterns can be rearranged to create different "colors" of music, called \textit{inversions} (illustrated in Figure \ref{fig:inversions} using the C major triad). Additionally, one of the first and most important things a new jazz pianist must learn is common chord progressions. Chord progressions are successions of chords that form the foundation of music \cite{levine11_jazzTheoryBook}. They are often represented using Roman numerals, with lowercase letters for minor chords and uppercase letters for major chords. This representation allows progressions to be independent of the specific key that a piece of music is played in. 

\subsection{Dataset creation via \textit{distance-based pattern structures (DBPS)}} \label{sec:creation}

\begin{algorithm}[t]
\small
\DontPrintSemicolon
\caption{Pattern generation via \textit{distance-based pattern structures (DBPS)}}\label{alg:chordGeneration}
\KwIn{distance = [$d_0$,$d_1$,$d_2$,...,$d_n$]; type="chord" or "arpeggio" or "scale"}
\KwOut{Pattern in all keys}
\ForEach{MIDI\_pitch \textbf{in} range(24, 109)}{
$note_0 \gets$ MIDI\_pitch\;

$time \gets time$ \textit{\#current time}

\For{i \textbf{in} range(0, len(distance))}{
$note_{i+1} \gets note_i + distance[i]$\;
\If{type == "arpeggio" or "scale"}{
$time \gets time+1$
}
}
}
\end{algorithm}

We use a method called \textit{distance-based pattern structures (DBPS)} to automatically generate piano patterns in a flexible and intuitive way. Thanks to the formulaic structure of most piano music patterns, the patterns can be generated based on the distance between the pitches within the patterns. We represent the pitches using MIDI pitch numbers, with a distance of 1 representing the shortest distance between two notes (i.e., a semitone). To illustrate this approach, we provide a simplified pseudocode in Algorithm \ref{alg:chordGeneration}. The algorithm takes the pitch distances from preceding notes as input to generate any pattern. For example, for any major triad with three notes $note_0, note_1, note_2$, the input would be $distance = [4,3]$. That is, given the base note as $note_0$, $note_1$ is 4 pitches (or 4 semitones) away, and $note_2$ is 3 pitches away. Similarly, for a maj7 tetrad, $distance = [4,3,4]$; and so on. 

To generate chord progressions, we use a similar approach based on the distances between notes within each chord of the progression. The base note (\textit{note0}) for each chord is defined by its distance from the reference note (the key in which the progression exists) based on the Roman numeral. For example, the base note of the $ii$ chord is two pitches from the reference note, while the base note of the $IV$ chord is five pitches away. Once the base note is determined, the chords in the progression can be generated as described in Algorithm \ref{alg:chordGeneration}. For example, the popular jazz chord progression, \textit{ii-V-I}, can be represented by the following pattern structure:

chord1: note0=ref+2; note1=note0+3; note2=note1+4

chord2: note0=ref+7; note1=note0+4; note2=note1+3

chord3: note0=ref; note1=note0+4; note2=note1+3

The DBPS approach can also be used to generate \textit{altered} or \textit{extended} chords commonly found in jazz music. The approach works by imposing the distance required for the alteration or extension to generate the new chord. For example, a dominant $7th$ extension on a chord with notes $note0, note1$, and $note2$, would add a new note $note3$, three pitches above $note2$. Adding a major $7th$ would add a note that is four pitches above $note2$. Similarly, alterations can be generated by modifying the pitch of a note within the chord. For instance, a $\flat5$ would be created by reducing the pitch of $note2$ by one.

We created pattern structures for all the piano patterns selected in our survey, and then used Python to develop a pattern generator (Section \ref{sec:patternGen}) that utilized these structures to generate MIDI format files\footnote{\url{https://github.com/MarkCWirt/MIDIUtil}} for all keys and inversions. These MIDI files were then manually reviewed for accuracy before being converted to WAV format using the Timidity synthesizer\footnote{\url{http://timidity.sourceforge.net/}}, with a sampling rate of 16KHz and 24-bit depth. To ensure the quality of the generated patterns, we randomly selected WAV files for verification by two jazz pianists, who played the patterns on physical pianos to confirm that the sounds and labels were correct. 

\subsection{Dataset description and statistics}

\tikzset{
  basic/.style  = {draw, text width=2cm, drop shadow, font=\sffamily, rectangle},
  root/.style   = {basic, rounded corners=2pt, thin, align=center, fill=white},
  level-2/.style = {basic, rounded corners=6pt, thin,align=center, fill=white, text width=1.6cm},
  level-3/.style = {basic, thin, align=center, fill=white, text width=1.3cm, text height=0.1cm}
}

\begin{figure}
\scriptsize
    \centering
\begin{tikzpicture}[
  level 1/.style={sibling distance=8em, level distance=3em},
%   {edge from parent fork down},
  edge from parent/.style={->,solid,black,thick,sloped,draw}, 
  edge from parent path={(\tikzparentnode.south) -- (\tikzchildnode.north)},
  >=latex, node distance=0.3cm, edge from parent fork down]

% root of the the initial tree, level 1
\node[root] {\textbf{jazznet dataset}}
% The first level, as children of the initial tree
  child {node[level-2] (c1) {\textbf{Chords}}}
  child {node[level-2] (c2) {\textbf{Arpeggios}}}
  child {node[level-2] (c3) {\textbf{Scales}}}
  child {node[level-2] (c4) {\textbf{Progressions}}};

% The second level, relatively positioned nodes
\begin{scope}[every node/.style={level-3}]
\node [below of = c1, xshift=5pt] (c11) {dyads};
\node [below of = c11, xshift=5pt, text width=1cm] (c111) {min2};
\node [below of = c111, text width=1cm] (c112) {maj2};
\node [below of = c112, text width=1cm] (c113) {min3};
\node [below of = c113, text width=1cm] (c114) {maj3};
\node [below of = c114, text width=1cm] (c115) {perf4};
\node [below of = c115, text width=1cm] (c116) {tritone};
\node [below of = c116, text width=1cm] (c117) {perf5};
\node [below of = c117, text width=1cm] (c118) {min6};
\node [below of = c118, text width=1cm] (c119) {maj6};
\node [below of = c119, text width=1cm] (c1110) {aug6};
\node [below of = c1110, text width=1cm] (c1111) {maj7\_2};
\node [below of = c1111, text width=1cm] (c1112) {octave};

\node [below of = c1112, xshift=-3pt] (c12) {triads};
\node [below of = c12, xshift=7pt, text width=1cm] (c121) {maj};
\node [below of = c121, text width=1cm] (c122) {min};
\node [below of = c122, text width=1cm] (c123) {aug};
\node [below of = c123, text width=1cm] (c124) {dim};
\node [below of = c124, text width=1cm] (c125) {sus2};
\node [below of = c125, text width=1cm] (c126) {sus4};

\node [below of = c126, xshift=-7pt] (c13) {tetrads};
\node [below of = c13, xshift=7pt, text width=1cm] (c131) {dim7};
\node [below of = c131, text width=1cm] (c132) {maj7};
\node [below of = c132, text width=1cm] (c133) {min7};
\node [below of = c133, text width=1cm] (c134) {min7b5};
\node [below of = c134, text width=1cm] (c135) {seventh};
\node [below of = c135, text width=1cm] (c136) {sixth};

\node [below of = c2, xshift=4pt] (c21) {dyads};
\node [below of = c21, xshift=5pt, text width=1cm] (c211) {min2};
\node [below of = c211, text width=1cm] (c212) {maj2};
\node [below of = c212, text width=1cm] (c213) {min3};
\node [below of = c213, text width=1cm] (c214) {maj3};
\node [below of = c214, text width=1cm] (c215) {perf4};
\node [below of = c215, text width=1cm] (c216) {tritone};
\node [below of = c216, text width=1cm] (c217) {perf5};
\node [below of = c217, text width=1cm] (c218) {min6};
\node [below of = c218, text width=1cm] (c219) {maj6};
\node [below of = c219, text width=1cm] (c2110) {aug6};
\node [below of = c2110, text width=1cm] (c2111) {maj7\_2};
\node [below of = c2111, text width=1cm] (c2112) {octave};

\node [below of = c2112, xshift=-6pt] (c22) {triads};
\node [below of = c22, xshift=8pt, text width=1cm] (c221) {maj};
\node [below of = c221, text width=1cm] (c222) {min};
\node [below of = c222, text width=1cm] (c223) {aug};
\node [below of = c223, text width=1cm] (c224) {dim};
\node [below of = c224, text width=1cm] (c225) {sus2};
\node [below of = c225, text width=1cm] (c226) {sus4};

\node [below of = c226, xshift=-7pt] (c23) {tetrads};
\node [below of = c23, xshift=8pt, text width=1cm] (c231) {dim7};
\node [below of = c231, text width=1cm] (c232) {maj7};
\node [below of = c232, text width=1cm] (c233) {min7};
\node [below of = c233, text width=1cm] (c234) {min7b5};
\node [below of = c234, text width=1cm] (c235) {seventh};
\node [below of = c235, text width=1cm] (c236) {sixth};

\node [below of = c3, xshift=7pt] (c31) {aeolian};
\node [below of = c31] (c32) {dorian};
\node [below of = c32] (c33) {ionian};
\node [below of = c33] (c34) {locrian};
\node [below of = c34] (c35) {lydian};
\node [below of = c35] (c36) {mixolydian};
\node [below of = c36] (c37) {phrygian};
\node [below of = c37] (c38) {pentatonic};

\node [below of = c4, xshift=5pt] (c41) {3-chord};
\node [below of = c41, xshift=11pt] (c411) {ii-V-I};
\node [below of = c411] (c412) {ii-V-i};
\node [below of = c412] (c413) {ii-triV-I};

\node [below of = c413, xshift=-11pt] (c42) {4-chord};
\node [below of = c42, xshift=11pt] (c421) {I-VI-ii-V};
\node [below of = c421] (c422) {i-vi-ii-V};
\node [below of = c422] (c423) {iii-VI-ii-V};
\node [below of = c423] (c424) {I-i$\sharp$-ii-V};
\node [below of = c424] (c425) {I-IV7-iii-VI7};
\node [below of = c425] (c426) {ii$\sharp$-V$\sharp$-ii-V};
\end{scope}
%\end{comment}
% lines from each level 1 node to every one of its "children"
\foreach \value in {1,...,12}
  \draw[->] (c11.195) |- (c11\value.west);
  
\foreach \value in {1,...,6}
  \draw[->] (c12.195) |- (c12\value.west);
  
\foreach \value in {1,...,6}
  \draw[->] (c13.195) |- (c13\value.west);
  
\foreach \value in {1,...,3}
  \draw[->] (c1.195) |- (c1\value.west);
  
\foreach \value in {1,...,12}
  \draw[->] (c21.195) |- (c21\value.west);
  
\foreach \value in {1,...,6}
  \draw[->] (c22.195) |- (c22\value.west);
  
\foreach \value in {1,...,6}
  \draw[->] (c23.195) |- (c23\value.west);
  
\foreach \value in {1,...,3}
  \draw[->] (c2.195) |- (c2\value.west);
  
\foreach \value in {1,...,8}
  \draw[->] (c3.195) |- (c3\value.west);
  
\foreach \value in {1,...,3}
  \draw[->] (c41.195) |- (c41\value.west);
  
\foreach \value in {1,...,6}
  \draw[->] (c42.195) |- (c42\value.west);
  
\foreach \value in {1,...,2}
  \draw[->] (c4.195) |- (c4\value.west);

\end{tikzpicture}
    \caption{High-level taxonomy of the jazznet Dataset.}
    \label{fig:taxonomy}
    \vspace{-10pt}
\end{figure}
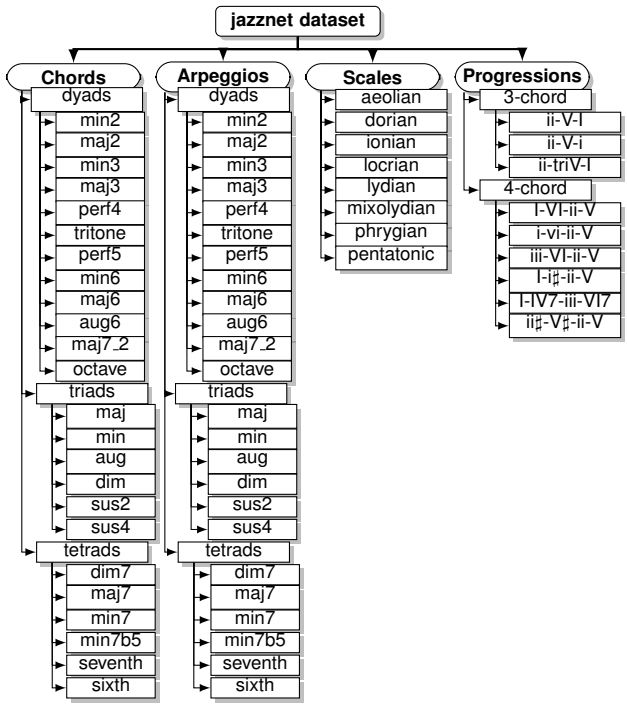

Figure \ref{fig:taxonomy} depicts a high-level taxonomy of the dataset. The dataset contains three kinds of chords and arpeggios: \textit{dyads}, \textit{triads}, and \textit{tetrads}. The scales are dominated by \textit{diatonic} (6-note) scales and include one \textit{pentatonic} (5-note) scale. The dataset is dominated by progressions, of which there are two kinds: 3-chord and 4-chord (tetrad) progressions. All the patterns are present in all their inversions and in all keys of the 88-key piano. 

Table \ref{tab:datasetStats} summarizes the dataset statistics. The dataset is published in WAV format (MIDI files are also provided) as separate data directories for \textit{types:} chords, arpeggios, scales, and progressions, and subdirectories for the different \textit{modes}. There are 65 labeled modes detailed as:

\begin{itemize}
    \item \textbf{Chords/Arpeggios} (24 each, appended with "-chord" or "-arpeggio"): \textit{12 dyads} (1 inversion each): minor 2nd (min2), major 2nd (maj2), minor 3rd (min3), major 3rd (maj3), perfect 4th (perf4), tritone, perfect 5th (perf4), minor 6th (min6), major 6th (maj6), minor 7th or augmented 6th (aug6), major 7th (maj7\_2), octave; \textit{6 triads} (2 inversions each): major (maj), minor (min), augmented (aug), diminished (dim), suspended 2nd (sus2), suspended 4th (sus4); and \textit{6 tetrads} (3 inversions each): dim7, maj7, min7, min7$\flat$5, seventh, sixth. There are 5,525 chords and 5,525 arpeggios in total.
    \vspace{-5pt}
    \item \textbf{Scales} (8): \textit{5 diatonics} (6 inversions each): aeolian, dorian, ionian, locrian, lydian, mixolydian, phrygian; \textit{1 pentatonic} (4 inversions). There are 4590 scales in total.
    \vspace{-5pt}
    \item \textbf{Progressions} (9): \textit{3-chord} (64 combinations): ii-V-I, ii-V-i, and ii-triV-I (triV means tritone substitution of V); \textit{4-chord} (256 combinations): I-VI-ii-V, i-vi-ii-V, iii-VI-ii-V, I-i$\sharp$-ii-V, I-IV7-iii-VI7, and ii$\sharp$-V$\sharp$-ii-V. There are 146,880 progressions in total.

\end{itemize}
\vspace{-5pt}
Each mode subdirectory contains the patterns in each mode labeled according to the specific note/pitch, octave, mode, and inversion. The inversions in the progressions are labeled according to the chord combinations.

The patterns are all recorded at a speed of 60 beats per minute (bpm), with two beats of decay at the end. The chord recordings are all 3 seconds long. The arpeggios and scales are recorded with one note played per beat. The arpeggios range from 4 to 6 seconds long, while the scales range from 7 to 9 seconds long, depending on the number of notes. Progressions are recorded with two chords played per measure and range from 7 to 10 seconds long. 
%, including the time and size of chords, arpeggios, scales, and progressions. %In total, the dataset contains 1,616,117$s$ (or 26,935 hours) of musical patterns.

\begin{table}[]
\scriptsize
\caption{Statistics of the jazznet dataset}
\label{tab:datasetStats}
\centering
\begin{tabular}{|l|l|l|l|l|l|l|l|}
\hline
Type      & \#modes & \#total & time (s) & \# hours & size (GB) \\ \hline
Chords    & 24          & 5,525    & 3        & 259  & 1            \\ \hline
Arpeggios & 24  & 5,525    & 4; 5; 6        &  433 & 1.7              \\ \hline
Scales    & 8  & 4,590    & 9; 7        & 674  & 6.3             \\ \hline
Progressions    & 9 & 146,880    & 7; 10        & 25,568   & 85.6         \\ \hline
\textbf{Total} & 65 & 162,520 &  & 26,934 & 95 \\ \hline
\end{tabular}
\vspace{-10pt}
\end{table}

\begin{comment}
\begin{table}[]
\tiny
\caption{Statistics of pattern types, modes, inversions, and time in the jazzNet dataset}
\label{tab:datasetStats}
\centering
\begin{tabular}{|l|l|l|l|l|l|l|l|}
\hline
Type      & \#modes & \#inversions & \#total & time (s) & \# hours & size (GB) \\ \hline
Chords    & 24          & 2; 3; 4         & 5,525    & 3        & 259  & 1            \\ \hline
Arpeggios & 24         & 2; 3; 4         & 5,525    & 4; 5; 6        &  433 & 1.7              \\ \hline
Scales    & 8        & 7; 5            & 4,590    & 9; 7        & 674  & 8.6             \\ \hline
Progressions    & 9        & 64; 256            & 146,880    & 14; 10        & 25,568   & 91.7         \\ \hline
\textbf{Total} & 65 & & 162,520 &  & 26,934 & 103 \\ \hline
\end{tabular}

\end{table}
\end{comment}

\subsection{Suggested subsets}

Given the dataset's imbalance resulting from the size of the chord progressions, we suggest some subsets of the dataset where a more balanced dataset may be required or downloading/using the entire dataset may be impractical (e.g., for initially testing a model). All subsets contain all the chords, scales, and arpeggios (15640 clips). The small, medium, and large subsets add 5876, 14688, and 36720 progressions, for a total of 21516, 30328, and 52360 clips, respectively. All the subsets are pseudo-randomly generated to ensure that all the modes are represented. The metadata files are in CSV format with a suggested split into train, validation, and test sets using an 80/10/10\% split. The validation and test sets are randomly selected from the different modes without overlapping with the training set. 

\subsection{Pattern generator} \label{sec:patternGen}
We have developed piano \textit{pattern generators} using Python, which allow users to easily create new patterns with ease. Due to space limitations, we omit the full details of the generator functions (the details are on the Github page\footnote{\url{https://github.com/tosiron/jazznet/tree/main/PatternGenerator}}), but summarize that chords, arpeggios, and scales can be generated by specifying the pitch distances as outlined in Section \ref{sec:creation}. In the same way, chord progressions can be generated by specifying the Roman numerals of the chords and the supported extensions/alterations (e.g., maj7, $\flat$5, $\sharp$, etc.). Crucially, the patterns are generated in all keys of an 88-key piano and in all inversions. The generated MIDI files can then be converted into the user's format of choice, such as WAV, using the user's choice of tool. The provided tool allows for the creation of numerous new patterns, and the scripts are open-source, allowing for the possibility of expanding the tool to support additional patterns.

%\subsection{Limitations of the dataset} \label{sec:limitations}

%The jazznet dataset has some limitations that may be overcome by other existing datasets. While these limitations do not detract from the dataset's overall utility, they are worth noting to provide context for the dataset's applications. Some of jazznet's limitations include the lack of dynamics (subjective interpretation in musical performance), which is contained in NSynth \cite{engel17_nsynth} in limited quantities. Jazznet also does not contain other features of jazz music like rhythm, tempo variances, etc, some of which may be present in other datasets, like FMA \cite{defferrard16_fmaDataset}. In addition, jazznet contains musical patterns for only one instrument, whereas other datasets like NSynth contain data for other musical instruments, albeit only for single notes. However, we anticipate that a similar approach to that described herein can be used to generate musical patterns for other stringed musical instruments. 
\vspace{-7pt}
\subsection{Applications}

Jazznet allows the evaluation of models developed for a variety of machine learning MIR tasks. The most straightforward and basic tasks involve \textbf{automatic music understanding} based on the dataset's class hierarchy: type recognition (whether an input is a chord, arpeggio, scale, or chord progression) or mode recognition, e.g., whether an input is augmented (aug) or major (maj) or Ionian or a ii-V-I progression. Specific details of the patterns can also be predicted, like the pitch contents, specific key (e.g., C-maj vs. E$\flat$-maj chord), octaves, and inversions. 

Jazznet can be used for more complex MIR tasks. For example, \textbf{automatic music transcription} \cite{benetos18_musicTranscription,benetos13_automaticMusicTranscription} is considered a canonical task in MIR, where the musical components (e.g., chords) in an audio recording are extracted. Most current music transcription models focus on isolated notes or chords present in the music. But additional musical patterns like scales and chord progressions can be predicted using models trained on jazznet. The dataset can also be used for the development of \textbf{music recommendation systems} \cite{schedl15_musicRecommender,van13_musicRecommendation}. Recommendation systems can be built based on musical patterns: a listener who likes certain kinds of chords or scales in one song (e.g., a song dominated by the minor scale or minor chords) may similarly like other songs that contain similar chord modes. Related to music understanding, jazznet can be used for \textbf{automatic music generation} \cite{herremans17_music_generation,carnovalini19_multilayeredMusicGeneration}. A lot of jazz music involves the repetition of jazz chord progressions with scales or arpeggios played over them. For example, the \textit{ii-V-I} progression, which appears in nearly all jazz standards \cite{levine11_jazzPianoBook}, can be soloed over with each chord's arpeggios or with matching scales (e.g., the Dorian mode over the \textit{ii} chord). Jazznet can be used to learn the fundamental patterns in musical pieces and how they occur (e.g., frequency of progressions, scales used with the progressions, etc.). %These patterns can then be used to generate new music pieces with some randomness involved, since improvisation also features a lot of randomness inserted in the patterns. We hope that the MIR research community will find the dataset useful for these and other machine learning applications.

\section{Experiments} \label{sec:experiments}
\begin{table}[]
\caption{Average precision and mAP for mode prediction.}
\label{tab:results}
\centering
\scriptsize
\begin{tabular}{|l|l|l|l|l|l|l|l|l|}
\hline
      & Chords & Arpeggios & Scales & Progressions & mAP \\ \hline
\textbf{CRNN} & 0.28    & 0.16 & 0.10   & 0.67 & 0.63           \\ \hline
\textbf{M5} & 0.36 &  0.09  & 0.06 & 0.48 & 0.30         \\ \hline

\end{tabular}
\vspace{-10pt}
\end{table}
\begin{comment}
\begin{figure}[t]
		\centering
		\includegraphics[width=0.7\linewidth]{figures/results.png}
		\vspace{-15pt}
		\caption{mAP of the CRNN and M5 models on the large subset}
		\label{fig:results}
       \vspace{-5pt}
\end{figure}
\end{comment}

\noindent\textbf{Experimental setup:} We used two simple models to demonstrate the potential of the jazznet dataset for music audio recognition in two tasks, type and mode recognition. The first model is a convolutional recurrent neural network (CRNN), as described in \cite{choi17_CRNN_music}. This model utilizes Mel-spectrograms as input and models long-term dependencies in musical structures using a recurrent neural network (RNN). The second model is the M5 model described in \cite{dai17_dnn_audio}: a deep convolutional neural network that takes time series waveforms as input and requires no preprocessing. Since data preprocessing is often a bottleneck for audio ML, this model provides a sense of the performance achieved using the dataset without any preprocessing. Both models contain 4 convolutional layers with batch normalization and the ReLU activation function. We trained the models with the Adam optimizer and categorical crossentropy loss function, and evaluated their performance using the mean Average Precision (mAP) and AUC (Area Under the Curve) score, following prior work \cite{dai17_dnn_audio}, on the medium subset (30328 clips) of the dataset. 

\vspace{3pt}
\noindent\textbf{Results:} Identifying chords, arpeggios, scales, and progressions was a simple task for the models. The CRNN and M5 models had high scores with an average AUC score of 0.984. They consistently recognized all four types, with mAP scores of 0.99 and 0.97. In comparison, two jazz pianists informally attempted to recognize a random sample of the types and both achieved 100\% accuracy. 

On the other hand, recognizing modes across the 65 modes was more difficult for the models, as summarized in Table \ref{tab:results}. The CRNN and M5 models achieved mAP scores of 0.63 and 0.30, respectively, with an average AUC score of 0.04. Both models found the arpeggios and scales to be the most challenging with AP of 0.16 and 0.10 for CRNN and 0.09 and 0.06 for M5. However, the models performed best on the progressions with an AP of 0.67 and 0.48 for CRNN and M5, respectively. The humans also found mode prediction more difficult than type recognition, achieving an accuracy range of 57\% to 73\%. In summary, the dataset was designed to reflect the hierarchy of difficulty in recognizing musical patterns and provides an opportunity to benchmark models for challenging MIR tasks.

\section{Conclusion}

In this paper, we presented a \textit{jazznet}, a dataset of essential jazz piano music patterns and an open-source pattern generator, enabling researchers to benchmark machine learning models for complex music information retrieval (MIR) tasks. Our aim is for this dataset to contribute to advancing machine learning research in MIR. In future work, we plan to expand the dataset by including more musical auditory attributes such as dynamics and rhythmic variations, as well as more complex patterns like 5-note chords and longer chord progressions. Additionally, we intend to explore more sophisticated models to improve performance on tasks using the dataset and also further investigate the potential of the DBPS approach for automatically generating different kinds of data. 

\bibliographystyle{IEEEbib}
\bibliography{refs}

\end{document}